\documentstyle[aps,multicol,epsf,prb]{revtex}
\begin{document}
\draft
\title{Magnetic-field dependence of dynamical vortex response in
  two-dimensional Josephson junction arrays and superconducting films}
\author {Beom Jun Kim and  Petter Minnhagen}
\address {Department of Theoretical Physics,
Ume{\aa} University,
901 87 Ume{\aa}, Sweden}
\preprint{\today}
\maketitle
\begin{abstract}
The dynamical vortex response of a two-dimensional 
array of the resistively shunted Josephson junctions in a 
perpendicular magnetic field is inferred from simulations.
It is found that, as the magnetic field is increased
at a fixed temperature, the response crosses over from normal to
anomalous, and that this crossover can be characterized by 
a single dimensionless parameter. 
It is described how this crossover should be reflected in
measurements of the complex impedance for Josephson junction arrays and
superconducting films.
\end{abstract}

\pacs{PACS numbers: 74.40.+k, 74.60.Ge, 74.76.-w, 75.40.Gb}

\begin{multicols}{2}
Two-dimensional (2D) vortex physics is strongly reflected in
the properties of Josephson
junction arrays (JJA) and superconducting films, as well as 
high-$T_c$ superconductors.~\cite{minnhagen_rev,newrock_rev}
In the absence of a magnetic field, the phase transition is of the
Kosterlitz-Thouless (KT) type and is driven by the unbinding of
thermally created vortex-antivortex pairs. In the
presence of a perpendicular magnetic field, ordered flux lattices are
formed which melt at high enough temperatures. 
Although these thermodynamic aspects are well known,~\cite{minnhagen_rev,newrock_rev}
the dynamical response properties, which contain key information of the vortex physics 
and can be extracted from measurements of the complex impedance and the flux 
noise,~\cite{minnhagen_rev,newrock_rev,beom_flux}
are much less well understood.~\cite{minnhagen_rev,newrock_rev}
In the present paper we focus
on the temperature region above the 2D flux lattice melting and infer 
that the dynamical response as a function of the perpendicular
magnetic field has intriguing characteristics.

The features of vortex dynamics are contained
in the complex impedance $Z(\omega )$ or equivalently the
conductivity $\sigma(\omega)=1/Z(\omega)$, which can be expressed as
$\sigma(\omega)=1/R_N-1/i\omega L_k\epsilon(\omega)$ with the normal-state 
resistance $R_N$, the kinetic inductance $L_k$, and  the vortex
dielectric function $1/\epsilon(\omega)$ describing the effect of the 
vortices on the dynamics.~\cite{minnhagen_rev,ahns,beom_big}
In the present work we obtain $1/\epsilon(\omega)$ and the resistance
$R={\rm Re}[\sigma^{-1}(\omega=0)]$ from simulations and characterize
the vortex dynamics in terms of these quantities.

\begin{figure}
\centerline{\epsfxsize=7cm \epsfbox{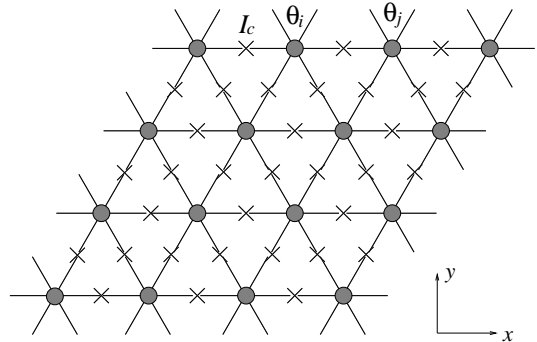}}
\vskip 0.5cm
\caption{ Triangular array of Josephson junctions. 
The superconducting island at site $i$ is associated with an order
parameter phase angle $\theta_i$ and the Josephson 
junction between adjacent islands with a critical current $I_c$.  
}
\label{fig_tri}
\end{figure}

We use the 2D resistively shunted junction (RSJ) model 
on an $L\times L$ triangular lattice with periodic boundary conditions 
(see Fig.~\ref{fig_tri}).
The equations of motion  in the presence of an external magnetic field
${\bf B} = B {\hat {\bf z}}$ follow from the local current conservation
\begin{equation} \label{eq_motion} 
\dot\theta_i = -\sum_j G_{ij} {\sum_k}^{'}\left[\sin(\theta_j-\theta_k-A_{jk}) + \eta_{jk}\right], 
\end{equation} 
where $\theta_i$ is the phase of the complex order parameter at site $i$,
$G_{ij}$ is the lattice Green's function,  and the primed summation is over six 
nearest neighbors of $j$. 
The thermal noise current $\eta_{jk}$ in units of the single junction critical current 
$I_c$  satisfies $\langle \eta_{ij}(t) \rangle = 0$ and 
$\langle \eta_{ij}(t) \eta_{kl} (0)\rangle = 2T(\delta_{ik}\delta_{jl} -  
\delta_{il}\delta_{jk})\delta(t)$, where $\langle \cdots \rangle$ is the ensemble 
average and the temperature $T$ is in units of the Josephson coupling strength 
$J \equiv \hbar I_c/2e$.
The magnetic bond angle $A_{ij} \equiv (2\pi / \Phi_0)\int_i^j{\bf A}\cdot d{\bf l}$
with the flux quantum $\Phi_0$ and the magnetic vector potential 
${\bf A} = B x {\hat{\bf y}}$ satisfies the
plaquette sum $\sum_p A_{ij} = 2\pi f$ with the frustration $f$, corresponding
to the number of flux quanta per elementary triangle.
The time  $t$ is in units of $\hbar/2eR_NI_c$ with the shunt resistance $R_N$, and 
in the following we simplify unit system to $R_N  = I_c = \hbar/2e = 1$.
We use a second-order algorithm for numerical integration with the time step
$\Delta t = 0.05$ and the system size $L=64$.
The key quantity is the time correlation function $G(t)$, 
which is related with the dielectric function $1/\epsilon(\omega)$ by 
\begin{equation} \label{G}
\frac{1}{i\omega} \left[ \frac{1}{\epsilon(\omega)} - \frac{1}{\epsilon(0)} \right] =
-\int_0^\infty dt e^{i\omega t}G(t),
\end{equation}
and is determined from 
$G(t)=(2\rho_0 T / \sqrt{3} L^2)\langle F(t)F(0)\rangle$,~\cite{houlrik,jonsson1,jonsson2}
where $\rho_0= \sqrt{3} \langle \cos(\theta_i-\theta_j-A_{ij})\rangle$ is the
superfluid density,~\cite{jonsson2,olsson} and 
$F(t) =  \sum_{\langle
  ij\rangle}\sin(\theta_i-\theta_j-A_{ij}){\bf e}_{ij}\cdot{\hat {\bf x} }$,
with the sum over all links and ${\bf e}_{ij}$ being the unit vector from $i$ to $j$.
In our unit system $\rho_0$ is
the same as the kinetic inductance $1/L_k$ so that
the relation between $\sigma(\omega)$ and
$1/\epsilon(\omega)$ reduces to
\begin{equation} \label{sigma}
  \sigma(\omega)=1-\frac{\rho_0}{i\omega\epsilon(\omega)}.
\end{equation}

\begin{figure}
\centerline{\epsfxsize=9cm \epsfbox[70 270 504 630]{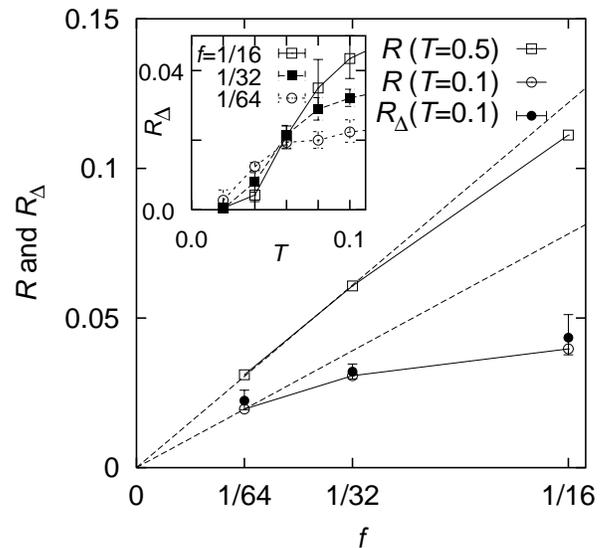}}
\vskip 0.0cm
  \caption{ Resistance $R$ as a function of frustration
    $f$ at $T=0.5$ (squares) and 0.1 (circles). The empty symbols represent $R$
    obtained as described in text, and the filled symbols as obtained
    by the method described in Ref.~\protect\onlinecite{beom_big} (denoted by $R_\Delta$).
    The dashed lines through the origin correspond to expectation from free vortex diffusion.
    The full drawn lines are guides to the eyes. The inset shows that the
    onset of resistance starts well below $T=0.1$.}
\label{fig_R}
\end{figure}

Figure~\ref{fig_R} shows the resistance $R$ as a function of frustration $f$ 
at $T=0.1$ and 0.5 obtained from $G(t)$ using Eqs.~(\ref{G}) and (\ref{sigma}) 
together with $R={\rm Re}[1/\sigma(\omega=0)]$. In order to check 
the present method we have also calculated the resistance 
in a completely independent way using the
fluctuating twist boundary conditions described in Ref.~\onlinecite{beom_big}. 
As seen in Fig.~\ref{fig_R} both determinations  give consistent
 results but the present method appears to converge much better. 
The inset in Fig.~\ref{fig_R} shows $R$ versus $T$ and verifies that 
$T=0.1$ and $0.5$  are both well above the resistive transition temperature.
The naive expectation is that $R$ is proportional to the density of free vortices which in
turn is proportional to $f$.~\cite{minnhagen_rev} This corresponds to a situation
where all vortices effectively move diffusively as independent particles. As
seen in Fig.~\ref{fig_R} this expectation is borne out for $T=0.5$ for the two
lowest frustrations $f=1/64$ and 1/32 (these data points fall on a straight line through
the origin). However, the highest frustration $f=1/16$ falls slightly below this line
indicating a suppression of the free vortex diffusion. This effect is much
more pronounced for the lower temperature $T=0.1$, where both 
resistances for $f=1/16$ and $f=1/32$ fall well below the straight
line through the origin. 
Since the vortex pinning caused by a triangular array is very weak,~\cite{lobb}
the essential part of the suppression is caused by some vortex correlation mechanism and
not by a lattice pinning effect.

\begin{figure}
\centerline{\epsfxsize=10cm \epsfbox[100 50 500 770]{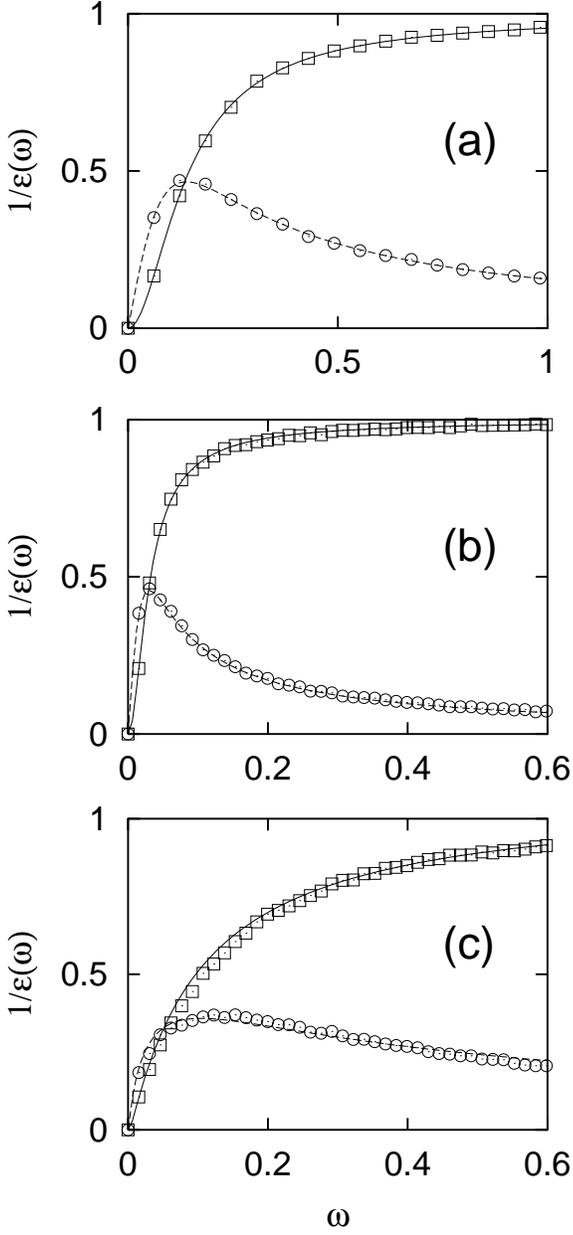}}
  \caption{The response function $1/\epsilon(\omega)$ for (a) $T=0.5$
    and $f=1/16$, (b) $T=0.1$ and $f=1/64$, and (c) $T=0.1$ and $f=1/16$.
    The full and dashed lines are the fits to the functional
    form described in text. For each pair $(T,f)$
this essentially gives a one-parameter characterization.
}
\label{fig_e}
\end{figure}

Figure~\ref{fig_e} shows the real and imaginary part of $1/\epsilon(\omega)$
obtained from $G(t)$ by aid of Eq.~(\ref{G}).
One immediately notes a qualitative difference; in Figs.~\ref{fig_e}(a) and (b) the
maximum of $|{\rm Im}[1/\epsilon(\omega)]$ is very close to the crossing point with
${\rm Re}[1/\epsilon(\omega)]$, whereas in Fig.~\ref{fig_e}(c)
this maximum is  displaced to the right.
In order to analyze these differences in a systematic way we parameterize the curves
by an interpolation between two response forms.
The first is the Drude form which corresponds to vortices
effectively diffusing as independent particles. This form is hence
expected to apply to the region where the resistance is
linear in $f$ (as for $f=1/64$ and $1/32$ at $T=0.5$ in Fig.~\ref{fig_R}).
The time correlation function in this case is of the exponential form
$G^{\rm Drude}(t)=(2/\tilde{\epsilon}\pi)e^{-t\sigma_0}$ 
and the corresponding response function is given by
\[ {\rm Re}\left[\frac{1}{\epsilon(\omega)}\right]=\frac{1}{\tilde{\epsilon}}
\frac{\omega^2}{\omega^2+\sigma_0^2}, \]
where $\tilde{\epsilon}$ describes an effective static polarization of
the vortex interaction due to vortex-antivortex  pairs,~\cite{minnhagen_rev}
and ${\rm Re}[1/\epsilon(\omega)]\propto \omega^2$ for small $\omega$.
The second response form describes the dynamical response of {\it bound} 
vortex-antivortex pairs whose phenomenological form has been proposed in 
Ref.~\onlinecite{minnhagen_rev}, Minnhagen phenomenology (MP),
and is given by Ref.~\onlinecite{beom_big}:
$ G^{\rm MP}(t)=(2/\tilde{\epsilon}\pi)[{\rm Ci}(\omega_0t)\sin\omega_0t-
{\rm si}(\omega_0 t)\cos \omega_0t]$, where ${\rm Ci}(x)\equiv -\int_x^\infty dt \cos t/t$, 
${\rm si}(x)\equiv-\int_x^\infty dt \sin t/t$, and $G^{\rm MP}\propto 1/t$ for large times.
The corresponding response function is given by
\[
{\rm Re}\left[\frac{1}{\epsilon(\omega)}-\frac{1}{\epsilon(0)}\right]
=\frac{1}{\tilde{\epsilon}}\frac{\omega}{\omega+\omega_0}, 
\]
which goes to zero linearly with $\omega$ and gives a vanishing resistance.
It is sometimes referred to as an anomalous vortex dynamics since it exhibits a nonanalytic
divergence of the conductivity for small $\omega$: 
$\sigma(\omega)\propto -\ln \omega$.~\cite{theron} This form has been shown, from simulations
of various models of 2D vortex physics in the absence of magnetic field, to describe the
response of the low-temperature phase,~\cite{beom_big,jonsson1,holmlund} where 
the dynamics is dominated by bound vortex-antivortex pairs.~\cite{minnhagen_rev}

In general both bound pairs and free vortices can exist  at the same
time, e.g., for $f=0$ above the KT transition and for a finite $f$ 
above the resistive transition, and the dynamical response is
expected to contain contributions from both of the response 
types.~\cite{beom_big,jonsson1} In the present work we use
the interpolation introduced in Ref.~\onlinecite{beom_big}:
$G^{\rm int}(t)=\tilde{\epsilon}G^{\rm Drude}(t)G^{\rm MP}(t)$,
which contains three parameters, $\tilde{\epsilon}$, $\sigma_0$, and $\omega_0$,
or equivalently $\tilde{\epsilon}$, $r=\omega_0/\sigma_0$, and $R$. 
Here $r$ is the important parameter which changes the response
from pure Drude for $r=0$ to pure MP for $r=\infty$, and $R$ 
satisfies  $\sigma_0=R[\rho_0(\pi +r\ln r)]/(1-R)[\pi\tilde{\epsilon}(1+r^2)]$ 
from $G^{\rm int}$ together with Eqs.~(\ref{G}) and (\ref{sigma}).
Since $R$ can be independently determined without curve fitting as in Fig.~\ref{fig_R},
we are left with only two fitting parameters $\tilde{\epsilon}$ and $r$. 
Furthermore, for a finite $f$ above the resistive transition there are
few bound vortex-antivortex pairs present in the static limit so that
$\tilde{\epsilon}\approx 1$,~\cite{minnhagen_rev} and the dynamical 
response is essentially characterized by the single parameter $r$.
In the following analysis we use $R$ in Fig.~\ref{fig_R}
and fit data to the dynamical response function $1/\epsilon(\omega)$ with
two free parameters $\tilde\epsilon$ and $r$. 
Figure~\ref{fig_e} shows three typical examples; the fits are very good and 
the obtained values $\tilde{\epsilon}$ are in
all cases close to 1 [ $\tilde{\epsilon}\approx 1.0$, $1.0$, and $0.93$
for Fig.~\ref{fig_e}(a), (b), and (c), respectively]. 

\begin{figure}
\centerline{\epsfxsize=8cm \epsfbox[100 280 453 630]{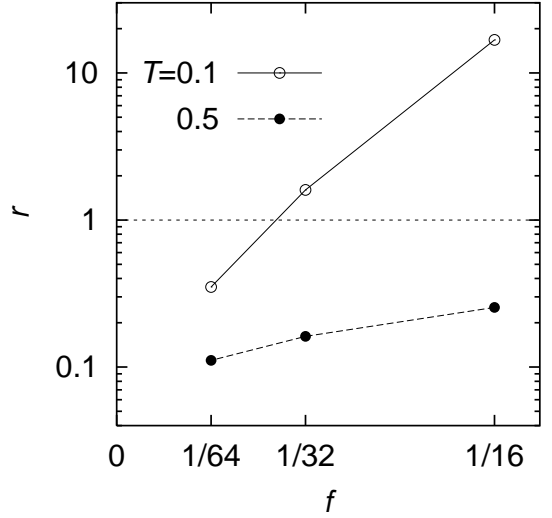}}
  \caption{The parameter $r$ characterizing the response as
    a function of $f$. At $T=0.5$ $r$ increases slightly with increasing $f$ 
    but remains much smaller than $1$, characterizing a Drude-like response.
    At $T=0.1$, $r$ changes from $r\ll 1$ to $r\gg 1$ with
    increasing $f$, signifying a crossover from Drude-like to MP-like response.
}

\label{fig_r}
\end{figure}

Figure~\ref{fig_r} shows the characterization of the dynamical response in terms
of the parameter $r$. The response for $T=0.5$ corresponds to a small $r$ and hence
Drude-like behavior. This is consistent with independent diffusion of {\it free}
vortices and with the linear relation between $R$ and $f$ found in
Fig.~\ref{fig_R}. However, as seen in Fig.~\ref{fig_R}, $r$ increases somewhat with $f$, 
which suggests an increase of ``vortex-pair-like'' correlations.
This seems consistent with the small suppression of $R$ with
increasing $f$ for $T=0.5$ in Fig.~\ref{fig_R}. The same feature is much more dramatic for
$T=0.1$: $r$ increases from a small Drude-like value to a large MP-like value with $f$, which
suggests that the response for the largest $f$ is dominated by
``vortex-pair-like'' correlations, and is consistent with the large suppression of $R$  
at large $f$ for $T=0.1$ in Fig.~\ref{fig_R}.

\begin{figure}
\centerline{\epsfxsize=8cm \epsfbox[100 280 453 630]{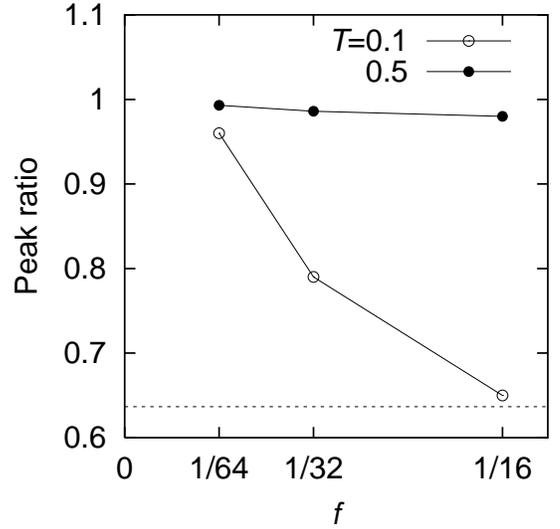}}
  \caption{Peak ratio as a function of frustration $f$. 
    At $T=0.5$ the peak ratio decreases slightly with increasing $f$
    but remains close to the Drude value $1$. At $T=0.1$ the peak ratio
    crosses over from $1$ to the MP value $2/\pi$.}

\label{fig_peak}
\end{figure}

An alternative way to analyze the crossover from Drude to MP response
is to focus on the peak ratio, defined as the ratio
$|{\rm Re}[1/\epsilon(\omega)]/{\rm Im}[1/\epsilon(\omega)]|$ at the maximum of
$|{\rm Im}[1/\epsilon(\omega)]|$.
For the pure Drude response this ratio is unity whereas for the pure MP response
it is $2/\pi\approx 0.64$ [see Fig.~\ref{fig_e} where (a) and (b) have peak ratios
close to the Drude value whereas the peak ratio in (c) is close to the MP value].
Figure~\ref{fig_peak} displays the peak ratios obtained directly from the data 
(by obtaining a sufficiently number of data points close to the maximum)
and shows that the crossover from Drude-like to MP-like behavior at $T=0.1$
with increasing $f$ is clearly signaled as a decrease in the peak ratio.

Our simulations show that in a certain parameter range the dynamical
response as a function of frustration crosses over from a free-particle-like
response to a response which
has the same form as the response of bound vortex-antivortex pairs
and that this crossover is
linked to a suppression of the resistance.
In Ref.~\onlinecite{jonsson2} the same type of crossover from Drude
to MP response with increasing frustration was found for a 2D $XY$ model
with time-dependent Ginzburg-Landau (TDGL) dynamics.
In Ref.~\onlinecite{theron}
it was found that a triangular JJA with a small finite frustration
shows a MP-like response well below the zero-frustration KT transition  and in
Ref.~\onlinecite{jonsson1} it was suggested, based on the peak ratio,
that the same is true for the superconducting film
measured in Ref.~\onlinecite{kapitulnik}.

Several theoretical attempts have been made to understand the origin
of the anomalous dynamics:~\cite{beck,capezalli1}
Ref.~\onlinecite{beck} ascribes the effect to a coupling between
spin waves and vortices, suggesting that there should be
anomalous dynamics for TDGL but not for RSJ, in contrast
to what has been demonstrated in the present work. 
Ref.~\onlinecite{capezalli1} attributes the effect to vortex 
correlations which should apply to both the TDGL and RSJ.
However, none of the theoretical formulations yields a crossover 
to the MP form and a shift of the peak ratio from 1 to $2/\pi$ with 
{\em increasing magnetic field} and hence with an {\em increasing resistance}. 
Thus an adequate theoretical explanation is still
largely lacking.
In Ref.~\onlinecite{jonsson2} it was found from simulations that an
increase of the magnetic field also causes an increased density of antivortices
suggesting that vortex-antivortex correlations may play a role
with a possible connection to 
the fact that the MP response also describes the response
of the low-temperature phase for $f=0$.

Our analyzing method makes it possible to assess to what extent the
crossover from normal to anomalous response described here shows up
in measurements on real systems.

The research was supported by the Swedish Natural Science Research Council
through Contract No. FU 04040-322.

\narrowtext

\end{multicols}

\end{document}